# Simulations of Jets Driven by Black Hole Rotation

Vladimir Semenov*, Sergey Dyadechkin* & Brian Punsly**

*Institute of Physics, State University St. Petersburg , 198504 Russia, **Boeing Space and Intelligence Systems, Torrance, California, USA

ABSTRACT:  The origin of jets emitted from black holes is not well understood, however there are two possible energy sources, the accretion disk or the rotating black hole. Magnetohydrodynamic simulations show a well-defined jet that extracts energy from a black hole. If plasma near the black hole is threaded by large-scale magnetic flux, it will rotate with respect to asymptotic infinity creating large magnetic stresses. These stresses are released as a relativistic jet at the expense of black hole rotational energy.  The physics of the jet initiation in the simulations is described by the theory of black hole gravitohydromagnetics.

Most quasars radiate a small fraction of their emission in the radio band (radio quiet), yet about 10% launch powerful radio jets (a highly collimated beam of energy and particles) with kinetic luminosities rivalling or sometimes exceeding the luminosity of the quasar host (radio loud).[1] There is currently no clear theoretical understanding of the physics that occasionally switches on these powerful beams of energy in quasars. Powerful extragalactic radio sources tend to be associated with large elliptical galaxy host that harbour supermassive black holes ($\sim 10^9$ $M_0$).[2] The synchrotron emitting jets are clearly highly magnetized and appe ar to emanate from the environs of the central black hole, within the resolution limits of very long baseline interferometry (VLBI). Observations (see the supplementary file) indicate that jets emitted from supermassive black hole magnetospheres are likely required for any theory to be in accord with the data.[2,3,4] Previous perfect magnetohydrodynamic (MHD hereafter) simulations of entire black hole magnetospheres have shown some suggestive results. For example, the simulations of  [5] were the first to show energy extraction from the black hole, but there was no outflow of plasma. Numerical models of an entire magnetosphere involve a complex set of equations that reflect the interaction of the background space-time with the plasma. In many instances, the only way to get any time evolution of the magnetosphere is to assume unphysical initial conditions. [5,6,9,10] Even so, previous simulations have not shown a black hole radiating away its potential energy into a pair of bipolar jets. [5,6,7,8,9,10] Most importantly, in previous efforts the underlying physics is masked by the complexity of the simulation. Thus, this numerical work has done little to clarify the fundamental physics that couples the jet to the black hole.

We exploit the simplification that the full set of MHD equations in curved spacetime indicate that a magnetized plasma can be regarded as a fluid composed of nonlinear strings in which the strings are mathematically equivalent to thin magnetic flux tubes. [11,12] In this treatment, a flux tube is thin by definition if the pressure variations across the flux tube is negligible compared to the total external pressure (gas plus magnetic), P,  that represents the effects of the enveloping magnetized plasma (the magnetosphere). By concentrating the calculation on individual flux tubes in a magnetosphere, we can focus the computational effort on the physical mechanism of jet production (on all the field lines).  Thus, we are able to elucidate the fundamental physics of black hole driven jets without burying the results in the effort to find the external pressure function, P. Our goal is to understand the first order physics of jet production not all of the dissipative second order effects that modify the efficiency and for these purposes the string depiction of MHD is suitable (see the methods section). The physical mechanism of jet production is a theoretical process known as gravitohydromagnetics (GHM) in which the rotating space-time geometry near the black hole drags plasma relative to the distant plasma in the same large-scale flux tube, thereby spring-loading the field lines with strong torsional stresses. [2]

The frame dragging potential of the rotating black hole geometry (described by the Kerr spacetime) is responsible for driving the jets. The frame dragging force is elucidated by the concept of the **minimum angular velocity** about the symmetry axis of the black hole as viewed from asymptotic infinity , $\Omega_P \geq \Omega_{min}$, where $\Omega_P$ is the angular velocity of the plasma (this frame is equivalent to Boyer-Lindquist coordinates, B-L hereafter). In flat space-time, $\Omega_P > -c/r\sin\theta \equiv \Omega_{min} < 0$, where c is the speed of light and $(r, \theta)$ are spherical coordinates. By contrast, within the ergosphere (located between the event horizon at $r_+$ and the stationary limit, where r, $\theta$, $\phi$,t are B-L coordinates), $\Omega_{min} > 0$. The ultimate manifestation of frame dragging is that all of the particle trajectories near the black hole corotate with the event horizon, $\Omega_{min} \to \Omega_H$ as $r \to r_+$ (**the horizon boundary condition**).[2] Now consider this $\Omega_{min}$ condition in a local physical frame at fixed poloidal coordinate, but rotating with $d\phi /dt \equiv \Omega$, so as to have zero angular momentum about the symmetry axis of the black hole, m = 0 (**the ZAMO frames**). In this frame (and in all physical frames i.e., those which move with a velocity less than c), a particle rotating at $\Omega_{min}$ appears to be rotating backwards, azimuthally relative to black hole rotation at c, $\Omega_P = \Omega_{min} \Rightarrow c\beta^\phi = -c$. From Eq. (3.49) of ref [2], the mechanical energy of a particle in B-L coordinates (i.e., the astrophysical rest frame of the quasar), $\omega < 0$ if $\beta^\phi < -c(r^2 - 2Mr + a^2)^{1/2}\sin\theta/(\Omega\ g_{\phi\phi})$, in particular:

$$\lim_{b^f \to 1} w = -m_\mu u^0 \frac{\Omega_{min}}{c}\sqrt{g_{ff}}, \ \Omega_{min} = \Omega - c(r^2 - 2Mr + a^2)^{1/2}\sin\theta/g_{\phi\phi} \ . \qquad (1)$$

In Eq. (1), the four-velocity of a particle or plasma in the ZAMO frames is given by $u^\lambda \equiv u^0(1, \mathbf{b})$, so $u^0$ is the Lorentz contraction factor that is familiar from special relativity, M is the mass of the black hole and "a" is the angular momentum per unit mass of the black hole in geometrized units, $\mu$ is the specific enthalpy, the B-L metric coefficient, $(g_{\phi\phi})^{1/2}$, is just the curved space analog of the azimuthal measure, $r\sin\theta$, of flat space-time up to a factor of a few. Since $\Omega_{min} > 0$ in the ergosphere, equation (1) implies that if a particle rotates so that $\Omega_P \approx \Omega_{min}$, then $\omega << 0$. Similarly, the specific angular momentum about the symmetry axis of the hole, $m = u^0 \beta^\phi (g_{\phi\phi})^{1/2}$, is negative for these trajectories as well. Hence, the in-fall of these particles toward the black hole is tantamount to extracting its rotational energy.

Our simulation in fig. 1 is of a perfect MHD (in terms of the field strength tensor, $F^{\lambda\nu}u_\nu = 0, \forall \lambda$) plasma, threading an accreting poloidal magnetic flux tube that begins aligned parallel to the black hole spin axis. The perfect MHD condition means that the plasma can short-out any electric fields that are generated in its rest frame. There are no existing numerical methods that can incorporate plasma dissipation into the black hole magnetosphere (i.e., go beyond perfect MHD), only the analytical work of ref [2] captures certain non-MHD effects and the results agree with our simulations. In order to choose a realistic initial state, note that ordered magnetic flux has been detected in the central 100 pc of some galaxies.[13] As Lens-Thirring torques concentrate the accreting plasma toward the plane orthogonal to the black hole spin axis, we expect the large scale flux to become preferentially aligned with the spin axis.[14] Long term numerical simulations of magnetized accretion show a concentration of vertical flux near the black hole.[15] Even an inefficient accretion of flux (i.e., reconnection of oppositely directed field lines and resistive diffusion) has been shown to lead to an enhanced flux distribution near the hole.[15,16] Consequently, we have chosen our simulations to follow the accretion of a poloidal magnetic flux tube that begins aligned parallel to the black hole spin axis. The consistent magnetospheric field configuration is a result of the long-term accumulation of similar flux tubes. A rapidly spinning black hole is chosen in the simulation (a/M=0.9998), as is often argued to be likely in a quasar.[17]

Initially, the flux tube is rotating with an angular velocity, $\Omega_F$, equal to the local ZAMO angular velocity, $\Omega_0$ - a rapidly decreasing function with radial coordinate.[2] Thus, initially, $\Omega_0 << \Omega_H$, because the flux tube is far from the event horizon. The flux tube accretes towards the black hole under the influence of the gravitational force. The natural state of plasma motion (geodesic motion) induced by frame dragging is to spiral

inwards faster and faster as the plasma approaches corotation with the event horizon

(**the horizon boundary condition**). By contrast, the natural state of plasma motion in a magnetic field is a helical Larmor orbit that is threaded onto the field lines. Generally,

these two natural states of motion are in conflict near a black hole. These two strong

opposing forces create a globally distributed torsion, creating the dynamical effect that

drives the simulation (fig. 1). The plasma far from the hole is still rotating slowly near

$\Omega_0$ in frame **a**. As the plasma penetrates the ergosphere, $\Omega_{min} \rightarrow \Omega_H$ as $r \rightarrow r_+$, and $\Omega_p$ must exceed $\Omega_0$ in short order since $\Omega_0 << \Omega_H$ (within $t \sim 0.1 GM/c^3$ after crossing the stationary limit). Thus, the ergospheric plasma gets dragged forward, azimuthally,

relative to the distance portions of the flux tube, by the gravitational field. The back

reaction of the field re-establishes the Larmor helical trajectories by torquing the plasma

back onto the field lines with **J** x **B** forces (the cross-field current density, **J**, driven by

this torsional stress is sunk within the enveloping magnetosphere). By Ampere's law, **J**

creates a negative azimuthal magnetic field, $B^\phi$, upstream of the current flow. The **J** x **B**

back reaction forces eventually torque the plasma onto $\omega < 0$ trajectories per Eq. (1).

Frames **b** and **c** illustrate the trend that the $B^\phi$ created in the ergosphere propagates upstream in the form of an MHD plasma wave at later times ($t > 70 \ GM/c^3$), as more and more negative energy (the red portion of the field line) is created in the ergospheric region of the flux tube. In frame **c**, a bonafide jet (a collimated, relativistic outflow of mass) emerges from the ergosphere.

Figure 1. Black Hole Driven Jet. A jet is produced on the magnetic flux tubes that experience the

torsional stress induced by the opposition of the gravitational frame dragging force with the **J** X **B**

electromagnetic force in which **J** is the current density and **B** is the magnetic field. The simulation is of a single flux tube in an enveloping magnetosphere of similar flux tube. The entire flux tube rotates in the same sense as the black hole, where $\Omega_H$ is the angular velocity of the event horizon as viewed from asymptotic infinity. The red portions of the field line indicate plasma with negative total energy, as viewed from asymptotic infinity. The black hole has a radius r $\approx 2GM/c^2$ which is $\approx 1.5 \times 10^{14}$ cm for a $10^9$ $M_0$ black hole. Frame **a** is a snapshot of the initiation of negative energy generation in the ergosphere: the outer boundary of the ergosphere, r $\approx GM/c^3(1+ \sin\theta)$. This effect is seen in ref [5], but that simulation ends at this early stage. In frame **b**, an outgoing Poynting flux emerges from the ergosphere and frame **c** shows a well-formed jet. The time lapse between frames, as measured by a distant stationary observer are: **a** to **b**, t = 78 $GM/c^3$ and **a** to **c**, t = 241 $GM/c^3 \approx 13$ days. The simulation ends with a pair of jets each over 50 $GM/c^2$ in length. Frame **d** is a closeup of the dynamo region in which $B^\phi$ is created in the jet, i.e. where $B^\phi$ changes sign. The pure Alfven speed (a measure of the ratio of magnetic energy to plasma inertia, $U_A = B^P/(4\pi\mu c^2)^{1/2}$) in the simulation is $U_A \approx 12c - 13c$ in the region in which Poynting flux is injected into the jet just above the dynamo.

The dynamo region for $B^\phi$ in the ergosphere is expanded in frame **d**. Because $B^\phi$ < 0 upstream of the dynamo, there is an energy ($\sim -\Omega_F B^\phi B^P$) and angular momentum ($\sim -B^\phi B^P$) flux along $B^P$, away from the hole in the jet. The red portion of the field line indicates that the total plasma energy per particle is negative, E < 0 (Since the magnetic field is primarily azimuthal near the black hole, $E \approx \omega + S/k$, where S is the poloidal Poynting flux along **B** and k is the poloidal particle flux along **B**) downstream of the dynamo. Because $B^\phi$ > 0 downstream, the field transports energy and angular momentum towards the hole with the inflowing plasma. Thus, S/k > 0 in the downstream state implying that $\omega$ << 0 in order for E < 0, downstream. The **J** x **B** back reaction forces on the twisted field lines torque the plasma onto trajectories with $\Omega_p \approx \Omega_{min}$. The ingoing $\omega$ << 0 plasma extracts the rotational energy of the hole by (1), thus black hole rotational inertia is powering the jet in the simulation. This is the fundamental physics of GHM.[2]

Three additional movies of simulations are provided with different initial conditions (movie 3 has the same P as fig.1, but the field line geometry is different; the flux tube is highly inclined, movie 4 has a different pressure function, $P \sim (r-r_+)^{-2.2}$ and movie 5 has four flux tubes) to show the generality of the results of the simulation in movies 1 and 2 (fig.1).

It is important to make a connection between what is modelled here and what is observed in quasar jets. The jet can be considered as bundle of thin flux tubes similar to those in our simulations (this is clearly visualized in movie 5). The jet is composed of Poynting flux and a relativistic outflow of particles. In the simulation of fig. 1, jet plasma attains a bulk flow Lorentz factor of 2.5 , at r ~ 50 GM/c$^2$ (which is ~ $10^{16}$ cm from a $10^9$ $M_0$ black hole which is assumed in all estimates to follow) in frame **c** ( at t ~ 250 GM/c$^3$). In quasars, VLBI observations indicate that quasar jet Lorentz factors are in the range of 2-30, with most ~ 10.[18] The resolution of VLBI is on the order 100 – 1,000 times the length of the jet in the simulations. Physically, it is believed that Poynting flux is required to accelerate the jet plasma to very high bulk flow Lorentz factors ~ 10 on parsec scales.[19,20] Furthermore, the magnetic tower created by B$^{\phi}$ in fig. 1, frame **c** in combination with the poloidal field component, B$^P$, naturally provides stable hoop stresses that are the only known collimation mechanism for the large scale jet morphology of quasars.[21,22] The nonrelativistic outer layer observed in some jets can be supported by a coexisting, enveloping relatively low power wind or jet from the accretion disk.[23]

The jets composed of a bundle of strings like those in the simulation are energetic enough to power quasar jets. The Poynting flux transported by a pair of bipolar collimated jets is approximately[2],

$$S \approx \frac{[\Omega_F \Phi]^2}{2\boldsymbol{p}^2 c} \, . \tag{2}$$

The total magnetic flux in the jet is $\Phi$. The field line angular velocity, $\Omega_F$, varies from near zero at the outer boundary of the ergosphere to the horizon angular velocity, $\Omega_H \approx 10^{-4} \sec^{-1}$, in the inner ergosphere. Thus, $S \sim (\Omega_H \Phi)^2 = (a\Phi/2Mr_+)^2 \sim (a\Phi/M)^2$ for rapidly rotating black holes. Consequently, there are three important parameters that determine jet power, M, a and B. Note that for rapidly spinning black holes the surface area of the equatorial plane in the ergosphere becomes quite large, for a/M = 0.996 it is $\approx 30\ M^2$.[2] Various accretion flow models yield a range of achievable ergospheric field strengths $B \sim 10^3\ G - 2\ x\ 10^4\ G$ that equate to $\Phi \sim 10^{33}\ G\text{-cm}^2 - 10^{34}\ G\text{-cm}^2$.[2,14,24] Inserting these results into equation (2) yields a jet luminosity of $\approx 10^{45} - 5\ x\ 10^{47}$ ergs/sec. This is consistent with the estimates of the kinetic luminosity of powerful quasar jets.[1] The maximum value of the flux noted above occurs when the persistent accretion of magnetic flux has a pressure that is capable of pushing the inner edge of the accretion disk out of the ergosphere (known as magnetically arrested accretion).[15,16] This maximum flux inserted in Eq. (2) equates to a jet power $\approx$ 5-25 times the bolometric thermal luminosity of the accretion flow in the disk models considered in refs [2, 24].

If 10% of the central black holes in quasars were magnetized by the accretion of vertical flux, this would explain the radio loud/radio quiet quasar dichotomy. To elevate this above a conjecture requires observational corroboration of the putative magnetosphere in radio loud quasars. A significant flux trapped between the black hole and the accretion disk should modify the innermost regions of the accretion flow. Thus, one can look for a distinction between radio loud and radio quiet quasar thermal emission at the highest frequencies. There might be evidence to support this already.

The accretion flow radiation has a high frequency tail in the EUV (extreme ultraviolet). Hubble Space Telescope data indicate that radio quiet quasars have an EUV excess compared to radio loud quasars.[25] The EUV suppression has been explained by the interaction of the magnetic field with the inner edge of the disk displacing the EUV emitting gas in radio loud quasars.[26]

The simulations presented here explain five important observations of radio loud quasars: a collimated jet is produced (based on radio observations), a power source (the black hole) that is decoupled from the accretion flow properties to first order (broadband, radio to UV, observations indicate that a quasar can emit most of its energy in a jet without disrupting the radiative signatures of the accretion flow, see the supplementary data), the suppression of the EUV in radio loud quasars (HST observations), the velocity of the jet is relativistic (VLBI data) and the maximal kinetic luminosity of the quasar jets (broadband radio and X-ray observations of radio lobes). The GHM process might also drive jets in other systems such as micro-quasars or gamma ray burst.[27,28] However, micro-quasars show correlations between accretion disk emission and jet properties, unlike quasars.[29] Consequently, there is no strong observational reason to prefer the black hole over the accretion disk as the primary power source as there is with quasar jets.

# Simulations of Jets Driven by Black Hole Rotation

Vladimir Semenov*, Sergey Dyadechkin* & Brian Punsly**

*Institute of Physics, State University St. Petersburg , 198504 Russia, **Boeing Space and Intelligence Systems, Torrance, California, USA

## SUPPLEMENTARY FILES

## The Independence of the Quasar Jet and the Accretion Flow: The case of 3C 82

The most compelling observational evidence for the existence of the GHM driven jets in quasars comes from optical/UV spectral data that indicates the independence of accretion flow luminosity with radio power to first order.  The radio powerful quasars have ultraviolet spectral signatures and luminosities (associated with the local viscous dissipation of the intense gas accretion towards the black hole) that are for the  most part indistinguishable from other quasars with feeble radio luminosity. Most of the small intrinsic contrasts in the emission line profiles are well described by differences in a distant "intermediate line region" that is not directly associated with  the accretion flow and is likely altered by jet propagation. [1,2] Intuitively, an accretion disk that describes a quasar like 3C 82 that has over ten times the kinetic luminosity in its jet compared to the broadband (IR to X-ray) luminosity due to viscous dissipation should have a spectrum significantly different than that of a typical radio quiet quasar, with about 0.1% of its power in a jet .However, the UV spectrum of 3C 82 is typical of a radio quiet quasar in terms of line profiles, spectral index and luminosity .[1,3] One could conjecture that the accretion disk model can extricate itself from this conundrum by launching the jet from a region that is distinct from the plasma that radiates the UV peak of the quasar emission. This does not seem likely since the far UV luminosity comes from the innermost regions of the accretion disk in virtually all models and in order to launch a strong relativistic jet one requires a sizeable surface area of the disk at these same small radii, where the plasma rotates the fastest.[4,5,6] The decoupling of UV spectral parameters from the radio jet power is very difficult to reconcile within the theory of accretion disk driven quasar jets.

The observational evidence presented above suggests that the black hole and not the accretion disk is the central engine of powerful quasar jets.

**The Case of  3C82**

3C 82 (also known as 4C43.09)  is a classical double radio source at z = 2.873 with 8.6 Jy of flux at 178 MHz which equates to $\approx$ 4 x $10^{40}$ W per the estimates developed in [7] and references therein. It has a moderate quasar luminosity $m_v$ =21.0.
The moderate 219 $\mu$ rest frame flux density of 1.08 ± 1.77mJy, from reprocessed accretion flow and star birth emission, indicates that the accretion flow luminosity is intrinsically moderate and not highly obscured. The kinetic luminosity of the jets is over ten times the accretion flow luminosity.

The far UV, 1000 Å- 2000 Å, spectral index of 3C 82 is 0.7 which is the average for radio quiet quasars of a similar bolometric luminosity.[3] The  UV broad-line properties are typical of a radio quiet quasar of similar luminosity.[1] In particular, in the quasar rest frame: CIV FWHM (full width half maximum) = 5,000 km/sec, EW (equivalent width) = 32 Å, Ly$\alpha$ FWHM = 4,500 km/sec, EW = 91 Å. Data on 3C 82 provided in collaboration with Steve Rawlings, private communication. The optical spectrum is provided below.

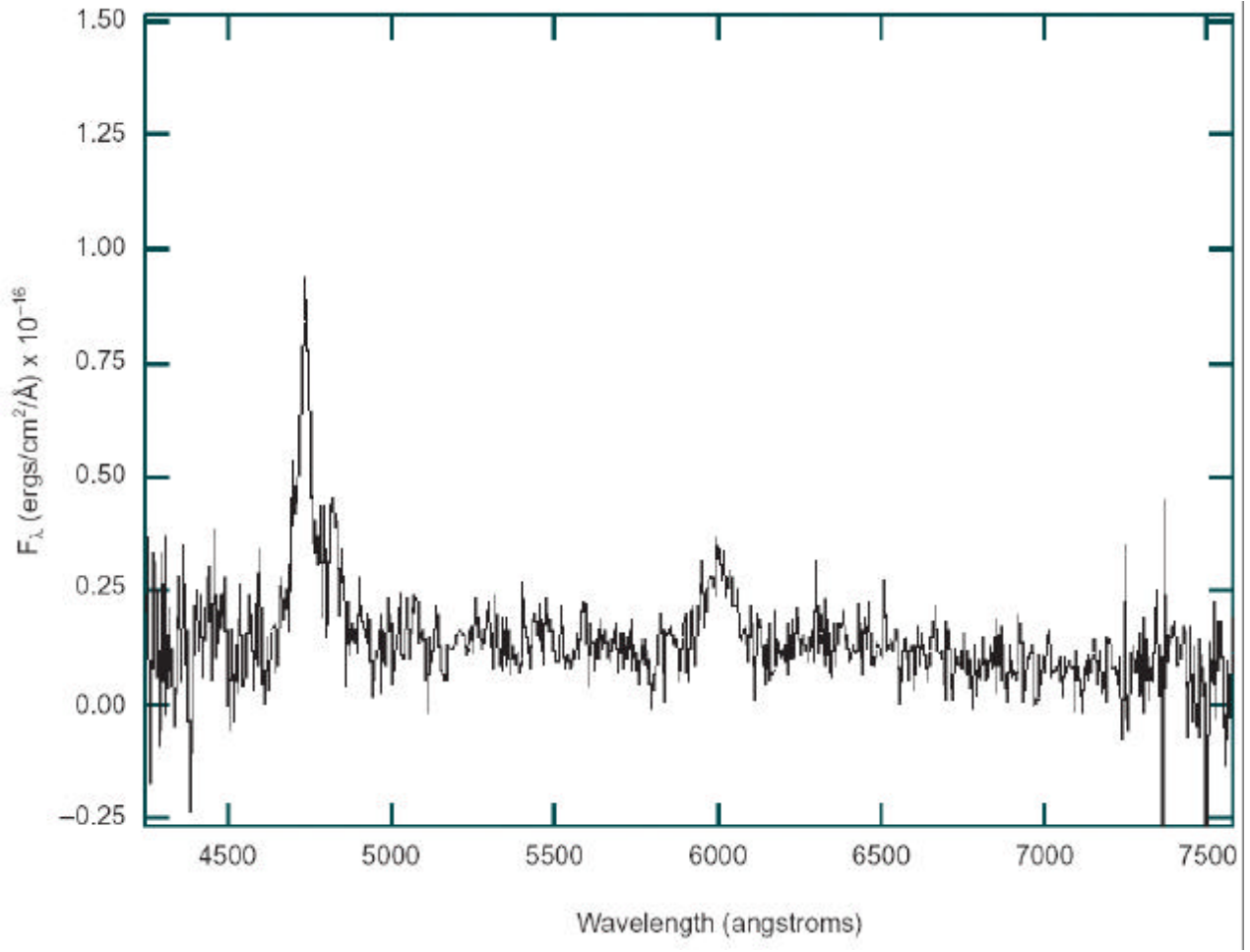

Figure S1. The optical spectra of 3C 82 . In the quasar rest frame the spectrum is in the UV and represents the peak of the thermal emission that is radiated from the accretion flow. The data reduction is courtesy of Steve Rawlings and we would like to thank Gary Hill for use of his telescope time on the McDonald Observatory 2.7 meter telescope.

# Simulations of Jets Driven by Black Hole Rotation.

# Methods.

Vladimir S. Semenov[1], Sergei A. Dyadechkin[1], and Brian Punsly[2]

[1]*Institute of Physics, State University, St. Petersburg, 198504 Russia,*

[2]*Boeing Space and Intelligence Systems, Torrance, California, USA*

## Abstract

We present here the necessary equations, description of the numerical code, and parameters of simulations so that others can reproduce our results. We also discuss the limitations and inherent assumptions of numerical methods.

## 1. Frozen-in coordinates

To derive the string equations, we have to introduce lagrangian co-ordinates into relativistic magnetohydrodynamic (RMHD) equations which can be presented as follows [1]:

$$\nabla_i \rho u^i = 0, \tag{0.1}$$

$$\nabla_i T^{ik} = 0, \tag{0.2}$$

$$\nabla_i(h^i u^k - h^k u^i) = 0. \tag{0.3}$$

Here equation (0.1) is the continuity equation, equation (0.2) is the energy-momentum equation, and equation (0.3) is Maxwell's equation; $u^i$ is the time-like vector of the 4-velocity, $u^i u_i = 1$,

$$h^i = *F^{ik} u_k \tag{0.4}$$

is the space-like 4-vector of the magnetic field, $h^i h_i < 0$, $*F^{ik}$ is the dual tensor of the electromagnetic field, and $T^{ij}$ is the stress-energy tensor:

$$T^{ij} = Qu^i u^j - Pg^{ij} - \frac{1}{4\pi} h^i h^j, \tag{0.5}$$

where

$$P \equiv p - \frac{1}{8\pi} h^k h_k, \ \ Q \equiv p + \varepsilon - \frac{1}{4\pi} h^k h_k. \tag{0.6}$$

Here $p$ is the plasma pressure, $P$ is the total (plasma plus magnetic) pressure, $\varepsilon$ is the internal energy including $\rho c^2$, and $g_{ik}$ is the metric tensor with signature $(1, -1, -1, -1)$.

The 4-vector of the magnetic field is orthogonal to the 4-velocity:

$$u_i h^i = 0, \tag{0.7}$$

which is evident because $*F^{ik}$ is an antisymmetric tensor.

We can introduce a coordinate system in such a way that the trajectory $u^i$ and magnetic field lines $h^i$ become coordinate lines. The fact is that a necessary and sufficient condition for two vectors $a^i$ and $b^i$ to act as coordinate lines is that they satisfy the following Lie equation [2,3]:

$$a^i \nabla_i b^k = b^i \nabla_i a^k. \tag{0.8}$$

Hence, we have to rewrite the frozen–in equation (0.3) in the form of the Lie equation (0.8). Generally speaking, $\nabla_i h^i \neq 0$, but we can find a function $q$ such that

$$\nabla_i qh^i = 0. \tag{0.9}$$

Then using (0.1) the Maxwell equation (0.3) can be rewritten in the form of a Lie derivative:

$$\frac{h^i}{\rho} \nabla_i \frac{u^k}{q} = \frac{u^i}{q} \nabla_i \frac{h^k}{\rho}, \tag{0.10}$$

and we can therefore introduce coordinates $\tau, \alpha$ such that:

$$x_\tau^i \equiv \frac{\partial x^i}{\partial \tau} = \frac{u^i}{q}, \quad x_\alpha^i \equiv \frac{\partial x^i}{\partial \alpha} = \frac{h^i}{\rho} \tag{0.11}$$

with new coordinate vectors $u^i/q, h^i/\rho$ tracing the trajectory of a fluid element and the magnetic field in a flux tube. This implies that we define a magnetic flux tube as a flux tube of vector field $h^i$, and a bundle of trajectories as a flux tube of vector field $u^i$. Bearing in mind the normalization $u^i u_i = 1$ it is clear that the function $q$ obeys:

$$q^2 g_{ik} x_\tau^i x_\tau^k = 1. \tag{0.12}$$

So, we have shown that ideal version of the induction equation (0.3) together with the continuity equation (0.1) allow us to introduce the Lagrangian coordinates $\tau, \alpha$ into the RMHD equations. The functions $x^i(\tau, \alpha)$ sweep the two dimensional manifold in the space-time continuum which consists of trajectories of the fluid elements for $\alpha = \text{const}$, and the magnetic field lines for $\tau = \text{const}$.

## 2. Variational method and string RMHD equations

Using (0.11) it is possible to derive the following relations:

$$\frac{u^i}{q} \nabla_i = \frac{\partial}{\partial \tau}, \tag{0.13}$$

$$\frac{h^i}{\rho} \nabla_i = \frac{\partial}{\partial \alpha}. \tag{0.14}$$

Then, the energy-momentum equation (0.2) can be rearranged to form a set of string equations in terms of the frozen-in coordinates:

$$\begin{aligned}
-\frac{\partial}{\partial \tau}\left(\frac{Qq}{\rho} x_\tau^l\right) - \frac{Qq}{\rho}\Gamma_{ik}^l x_\tau^i x_\tau^k \\
+ \frac{\partial}{\partial \alpha}\left(\frac{\rho}{4\pi q} x_\alpha^l\right) + \frac{\rho}{4\pi q}\Gamma_{ik}^l x_\alpha^i x_\alpha^k = -\frac{g^{il}}{\rho q}\frac{\partial P}{\partial x^i},
\end{aligned} \tag{0.15}$$

where $\Gamma_{ik}^l$ is the Christoffel symbol.

The string equations (0.15) for a flux tube embedded in a gravitational field $g_{ik}(x^i)$ and a pressure field $P(x^i)$ can also be derived from the action [4]:

$$S = -\int L(x^i_\tau, x^i_\alpha, x^i)d\tau d\alpha = -\int \frac{Q}{\rho}\sqrt{g_{ik}x^i_\tau x^k_\tau}d\tau d\alpha, \qquad (0.16)$$

where $L$ is the Lagrangian density.

The action (0.16) is invariant under $\tau$-reparametrization $\tau \to \tau'(\tau)$. Therefore the canonical Hamiltonian

$$\frac{\partial L}{\partial x^i_\tau}x^i_\tau - L, \qquad (0.17)$$

is easily seen to vanish identically. This implies that equations (0.15) are not independent, and we need a gauge condition to fix parametrization.

For simplicity we derive the gauge constraint for the particular case when entropy is initially uniform along the tube. Because entropy is conserved in the course of the motion [1], we can conclude that entropy does not depend on $\alpha$ for all $\tau$. Hence, the Lagrangian (0.16) does not depend explicitly on $\alpha$, which leads to the following equation:

$$\frac{\partial}{\partial \alpha}\left(\frac{\partial L}{\partial x^i_\alpha}x^i_\alpha - L\right) = -\frac{\partial}{\partial \tau}\left(\frac{\partial L}{\partial x^i_\tau}x^i_\alpha\right), \qquad (0.18)$$

and we find that:

$$\frac{\partial}{\partial \alpha}\left(w\sqrt{g_{ik}x^i_\tau x^k_\tau}\right) = \frac{\partial}{\partial \tau}\left(\frac{Q}{\sqrt{g_{ik}x^i_\tau x^k_\tau}}g_{ik}x^i_\tau x^i_\alpha\right). \qquad (0.19)$$

Here $w = \varepsilon + p/\rho$ is the enthalpy of the plasma.

Four vectors of velocity and magnetic field are orthogonal (0.7), hence the right hand side of the equation (0.19) has to vanish which immediately yields:

$$w\sqrt{g_{ik}x^i_\tau x^k_\tau} = f(\tau), \qquad (0.20)$$

where $f(\tau)$ is an arbitrary function. It is appropriate to choose parametrization $\tau \to \tau'(\tau)$ so that $f(\tau) = 1$, which corresponds to the following gauge condition:

$$q = \frac{1}{\sqrt{g_{ik}x^i_\tau x^k_\tau}} = w. \qquad (0.21)$$

Using the gauge condition (0.21) it can be shown that equations (0.15) are of hyperbolic type with relativistic alfvénic characteristics

$$\frac{d\alpha}{d\tau} = \pm \frac{\rho}{w\sqrt{4\pi Q}}, \tag{0.22}$$

and slow mode characteristics

$$\frac{d\alpha}{d\tau} = \pm \sqrt{\frac{\kappa \rho p}{4\pi w^3 (2P + p(\kappa - 2))}}. \tag{0.23}$$

By formulating the RMHD equations in terms of frozen-in coordinates, the energy-momentum equation (0.2) reduces to a set of 1D wave equations (0.15), which are in fact nonlinear string equations. The behavior of flux tubes can therefore be studied through solving a set of string equations, and an analogy can be drawn between the behavior of strings and flux tubes.

For a cyclic variable $x^m$ there is a conservation law which states that:

$$\frac{\partial}{\partial \tau}\left(\frac{\partial}{\partial x_\tau^m}(\frac{Q}{q\rho})\right) = -\frac{\partial}{\partial \alpha}\left(\frac{\partial}{\partial x_\alpha^m}(\frac{Q}{q\rho})\right). \tag{0.24}$$

The Kerr metric in Boyer-Lindquist coordinates (denoted by $g_{\mu\nu}$, hereafter) is given by [2]:

$$ds^2 = (1 - \frac{2Mr}{\Sigma})dt^2 - \frac{\Sigma}{\Delta}dr^2 - \Sigma d\theta^2$$
$$- (r^2 + a^2 + \frac{2Mra^2}{\Sigma}\sin^2\theta)\sin^2\theta\, d\varphi^2$$
$$+ \frac{4Mra}{\Sigma}\sin^2\theta\, d\varphi\, dt, \tag{0.25}$$

where

$$\Delta = r^2 - 2Mr + a^2, \;\; \Sigma = r^2 + a^2\cos^2\theta. \tag{0.26}$$

Here $M$ and $a$ are the mass and angular momentum of the hole, respectively, and we have used a system of units in which $c = 1, \;\; G = 1$.

Let us now consider a test flux tube which falls from infinity into a Kerr black hole. For cyclic variables $t$ and $\varphi$, the energy and angular momentum conservation laws for the flux tube can be written as:

$$\int_{\alpha_1}^{\alpha_2} \frac{Q}{w\rho}(g_{tt}t_\tau + g_{t\varphi}\varphi_\tau)d\alpha = E, \tag{0.27}$$

$$\int_{\alpha_1}^{\alpha_2} \frac{Q}{w\rho}(g_{t\varphi}t_\tau + g_{\varphi\varphi}\varphi_\tau)d\alpha = -L, \tag{0.28}$$

if there is no flux of energy and angular momentum through the ends $\alpha_1, \alpha_2$ of the flux tube.

## 6. Numerical method

**6.1 Normalization**. We normalize plasma density to its initial value $\rho_0$, length to the radius of the ergosphere in the equatorial plane $r_g$, velocity to the light speed $c$, time scale to the ratio $r_g/c$, magnetic field to $\sqrt{4\pi\rho_0}c$, the plasma pressure and the energy density to $\rho_0 c^2$, $\alpha$ to $r_g\sqrt{\rho_0/(4\pi c^2)}$. Note that radius of the event horizon is equal to $1/2$ for an extreme rotating black hole.

**6.2 Total pressure**. We choose for simulation the following distribution:

$$P(r) = \frac{c_P}{(r-r_g)^d}, \tag{0.29}$$

where $c_P$ and $d$ are some constants.

**6.3 String equations**. From the numerical point of view it is convenient to rewrite the string equations (0.15) in conservative form using the variational method:

$$\frac{\partial}{\partial\tau}\left(\frac{Qw}{\rho}x_{i\tau}\right) - \frac{\partial}{\partial\alpha}\left(\frac{\rho}{4\pi w}x_{i\alpha}\right) = \frac{1}{\rho w}\frac{\partial P}{\partial x^i}$$
$$+ \frac{1}{2}\frac{\partial g_{jk}}{\partial x^i}\left(\frac{Qw}{\rho}x_\tau^j x_\tau^k - \frac{\rho}{w}x_\alpha^j x_\alpha^k\right), \tag{0.30}$$

where $x_{i\tau} \equiv g_{ik}x_\tau^k, x_{i\alpha} \equiv g_{ik}x_\alpha^k$. The equations (0.30) have an advantage in comparison with (0.15) in that energy density (0.27) and angular momentum density (0.28) are explicitly included in the numerical scheme.

Adding the following equations:

$$\frac{\partial x_\alpha^i}{\partial\tau} = \frac{\partial x_\tau^i}{\partial\alpha}, \tag{0.31}$$

$$\frac{\partial x^i}{\partial\tau} = x_\tau^i, \tag{0.32}$$

we can present the string equations in terms of 12-dimensional vectors of state $W$, flux $F$ and source $S$:

$$\frac{\partial W}{\partial \tau} - \frac{\partial F}{\partial \alpha} = S, \qquad (0.33)$$

where

$$W = \left( \frac{Qw}{\rho} x_{i\tau}, x_\alpha^i, x^i \right), \qquad (0.34)$$

$$F = \left( \frac{\rho}{w} x_{i\alpha}, x_\tau^i, 0 \right), \qquad (0.35)$$

$$S = \left( \frac{1}{\rho w} \frac{\partial P}{\partial x^i} + \frac{1}{2} \frac{\partial g_{jk}}{\partial x^i} \left( \frac{Qw}{\rho} x_\tau^j x_\tau^k - \frac{\rho}{w} x_\alpha^j x_\alpha^k \right), 0, x_\tau^i \right). \qquad (0.36)$$

We solved the string equation (0.33) numerically using total variation diminishing (TVD) scheme [5]. To avoid short wavelength fluctuations inside the ergosphere three-point filter has been applied each third step so that $W_k$ has been replaced by $(W_{k-1} + 2W_k + W_{k+1})/4$. The filtering allows to reduce the errors a bit and perform calculations for longer time.

**6.4 Calculation of density**. The plasma is assumed to be polytropic with the following equations of state for gas pressure, internal energy and enthalpy in dimensionless form:

$$p = c_e \rho^\kappa, \quad \varepsilon = 1 + \frac{c_e}{\kappa - 1} \rho^{\kappa-1}, \quad w = 1 + \frac{c_e \kappa}{\kappa - 1} \rho^{\kappa-1}. \qquad (0.37)$$

Note that the relativistic term $\rho c^2$ is taken into account, and, for example, the enthalpy per unit mass is equal to $w = c^2 + \frac{p}{\rho}$.

The plasma density is obtained from the following transcendental equation:

$$P(x^i) = c_e \rho^\kappa - \frac{1}{2} \rho^2 g_{ik} x_\alpha^i x_\alpha^k, \qquad (0.38)$$

which is just a definition of the total pressure. For the particular case $\kappa = 2$ the density can be found analytically:

$$\rho = \sqrt{\frac{P(x^i)}{c_e - \frac{1}{2} g_{ik} x_\alpha^i x_\alpha^k}}. \qquad (0.39)$$

For the sake of simplicity we will suppose the polytropic index to be 2, although the results obtained are not sensitive to this value.

**6.5 Initial state**. For the initial moment $\tau = 0$ the flux tube is assumed to be elongated along the $x$-axis outside the ergosphere:

$$x^1 = c_x\alpha + x_0, \;\; x^2 = c_y\alpha + y_0, \;\; x^3 = c_z\alpha + z_0, \tag{0.40}$$

where $c_x, c_y, c_z, x_0, y_0, z_0$ are constants. Then the plasma density is determined from equation (0.39).

We assume that the initial angular momentum of each element of the string vanishes, hence

$$\phi_\tau = -\frac{g_{t\phi}}{g_{\phi\phi}}t_\tau. \tag{0.41}$$

Then the initial velocity $t_\tau$ can be obtained from the gauge condition (0.21)

$$t_\tau = \frac{1}{w}\sqrt{\frac{g_{\phi\phi}}{g_{tt}g_{\phi\phi} - g_{t\phi}^2}}. \tag{0.42}$$

Using equations (0.40- 0.42) we can find $W, F$ and $S$ for the initial moment $\tau = 0$.

**6.6 Boundary conditions**. The length of the flux tube must be sufficiently long to avoid influence of energy and momentum flux through the edges of the string. We used free boundary conditions at the edges:

$$\frac{\partial}{\partial\alpha}x_\tau^i = 0, \tag{0.43}$$

and we controlled conservation of the total energy and angular momentum of the string to within 1% of their initial values. The tube is sufficiently long so that disturbances from the edge do not influence the ergospheric process and disturbances from the ergosphere can not reach the ends of the tube for the duration of simulations. Hence boundary conditions are not really important.

The event horizon is not actually a boundary to the simulation. Due to gravitational redshifting, the tubes never cross or reach the event horizon [6]. Thus, there is no reason to excise the event horizon or impose boundary conditions.

**6.7 Tortoise variable**. To avoid too big a step along the radial coordinate $r$ we used the tortoise distance [4]

$$s = -\ln(r - r_h), \tag{0.44}$$

where $r_h$ is the radius of the event horizon.

**6.8 Error control**. The problem under consideration is a stiff one, and at some stage errors of calculation started to grow exponentially. Therefore it is important to control accuracy of the numerical scheme. To this end we used the gauge condition (0.21) which has not been actually involved in the numerical method. If the difference

$$w\sqrt{g_{ik}x^i_\tau x^k_\tau} - 1 < \varepsilon_0 \tag{0.45}$$

exceeds some level $\varepsilon_0$ we stop the calculations. The value $\varepsilon_0 = 0.1$ has been taken in most of the runs.

**6.9 Initial parameters**. We present results of our simulation for the following parameters:

1. Initial magnetic field is parallel to the spin axis of the hole (movies 1 and 2)

$$c_e = 4.0, \;\; c_P = 0.02, \;\; c_x = 0, \;\; c_y = 0, \;\; c_z = 1, \;\; x_0 = 1.5, \;\; y_0 = 1.5, \;\; z_0 = 0, \;\; d = 2. \tag{0.46}$$

2. Initial magnetic field is inclined to the spin axis of the hole (movie 3)

$$c_e = 4.0, \;\; c_P = 0.02, \;\; c_x = 0, \;\; c_y = 0.75, \;\; c_z = 0,75 \;\; x_0 = 1.5, \;\; y_0 = 1.5, \;\; z_0 = 0, \;\; d = 2. \tag{0.47}$$

3. Initial magnetic field is parallel to the spin axis of the hole (movie 3)

$$c_e = 4.0, \;\; c_P = 0.02, \;\; c_x = 0, \;\; c_y = 0, \;\; c_z = 1, \;\; x_0 = 1.5, \;\; y_0 = 1.5, \;\; z_0 = 0, \;\; d = 2. \tag{0.48}$$

For the four flux tube simulation (movie 5)the initial positions of the tubes were chosen symmetrically $x_0 = \pm 1.5, \;\; y_0 = \pm 1.5, \;\; z_0 = 0$ with the other parameters the same as in 1.

The number of grid points along the string $-20 < \alpha < 20$ is chosen to be 8000. The black hole is supposed to be near the extreme rotation $a/M = 0.9998$. For additional details see [4].

## 7. Limitations of the Method

A major shortcoming of this method is that the strings can cross the other field lines with no consequence within the simulation even though this is not physical. In reality, this should result in dissipative effects such as reconnection and local inhomogeneous pressure gradients. These effects could limit the efficiency of the jet to transport energy to large distances. The more accurately P is determined ahead of time, the less field line crossings occur during the accretion of the flux tube. For example, the geometry of the jet relative to the background magnetosphere in movie 3 is highly disordered and requires many field line crossings and is not likely to produce an efficient jet. However, for a uniform accretion of flux (such as the accretion of vertical field lines in movies 1,2 ,4 and 5) field line crossings do not occur and the dissipative effects will not be an issue (this is clear from movie 5 in which a series of flux tubes accrete one behind another with the same dynamics that is seen in the single flux tube simulations).

From an astrophysical point of view, a low plasma density is more physical. We have tried to get the simulations as magnetically dominated as possibly while still showing a long well-formed jet. In the simulations, the magnetic dominance is given by the pure Alfven speed (defined in the caption to fig.1), $U_A \approx 12c - 13c$ in the ergosphere just upstream of the dynamo. In powerful radio quasars, typical values are probably $10 < U_A < 10^6$ [6]. Error control becomes a major problem when the density becomes too low. As more plasma accretes toward the black hole and more plasma is ejected by the jet, the density gets lower and lower. This is a limiting factor for the length of our simulations. The gradients in the dynamical quantities get very large near the horizon making it difficult to control errors without a very a large number of points and huge computer memory. If the density is too low, the simulations advance only as far as frames a or b of fig.1 before the simulations crashes as a result of large errors. Negative energy creation is seen, but not a well-formed jet.

Eventually the density gets so low that pair creation processes become significant for loading the field lines. The most likely mechanism for plasma loading in a luminous quasar is pair creation from a gamma ray field that is the high frequency tail of the accretion disk coronal spectrum. The density that is required to establish perfect MHD (the Goldreich-Julian charge density) is $\sim 0.1\text{cm}^{-3}$ in the black hole magnetosphere see Chapter 9 of [6]. In Chapter 6 of [6], it is shown that if the MeV luminosity of the accretion disk corona is above $10^{41}$ ergs/sec, one expects this density to be exceeded by an order of magnitude and MHD flux tubes should be representative of the black hole magnetosphere (i.e., if at least 0.001 percent of the bolometric luminosity of the accretion flow is in the form of low energy gamma rays). The analytical models of [6] indicate that the same physical process of GHM launched jets found in our simulations still exists when $U_A \gg 10$ since the torsional stress on the large-scale field lines will still be created by the frame dragging force.

Another limitation of the method is the need for a rapidly rotating black hole, $a/M \approx 1$. For lower values of the spin parameter such as $a/M = 0.8$, one finds that the jet formation proceeds too slowly to control the errors. As more and more plasma is sucked very close to the horizon by the gravitational force, the gradients in dynamical quantities become very large. Simultaneously, the density becomes too low farther out in the ergosphere where much of the GHM interaction occurs, thus producing large gradients and errors in an extended region. Due to the large gradients, the simulations advance only as far as frames a or b of fig.1 before the simulations crash as a result of large errors. Negative energy creation is seen, but not a well-formed jet. However, as noted in the text this might be academic for quasars as the large accretion rates associated with the quasar emission and Lens-Thirring torques probably make $a/M \approx 1$ a common occurrence.

## 8. The Pressure Function

As discussed above, the more accurate choice of P, the more faithful the string simulation is to a space-filling MHD simulation. Thus, P should be chosen with care. The diverging pressure function, near the horizon, is magnetic in origin and is a consequence of the manner in which pressure is defined in the string formalism. The large magnetic contribution to the pressure results from a combination of the electromagnetic forces being on the same order as the inertial forces (gravity) and the infinite gravitational redshift at the horizon, making it appear that the field gets infinitely twisted azimuthally. The long term accretion of vertical magnetic flux discussed in the text is the magnetically arrested accretion condition of [7]. In this scenario, the magnetic field is dragged radial inward along the equatorial plane as seen in the simulation of [8]. The magnetic tension in the stretched poloidal magnetic field, $\mathbf{B^P}$, can be described as a force $\mathbf{J} \times \mathbf{B^P}$ that opposes the radial attraction of gravity (see chapter 8 of [6]). Thus, very close to the event horizon the inflow velocity as seen by the ZAMOs, $v_P$, can be significantly less than c. Consider the total magnetic pressure, $\frac{1}{2}F^{\mu\nu}F_{\nu\mu}$, from the MHD expression (5.35) of [6] evaluated in the relevant case that magnitude of the total velocity (azimuthal plus poloidal) in the ZAMO frame is approximately c near the horizon,

$$\frac{1}{2}F^{\mu\nu}F_{\nu\mu} \approx \left[1 + \frac{(\Omega_F - \Omega_H)^2 g_{\phi\phi}^4 (c^2 - v_P^2)}{\Delta^2 c^2 v_P^2}\right](B^P)^2 \,, \qquad (0.49)$$

where we used the definitions of Kerr metric quantities from Eqs. (0.24) and (0.25). The divergence in $\frac{1}{2}F^{\mu\nu}F_{\nu\mu}$ comes from the $\Delta^2$ term in the denominator. For magnetically arrested accretion and $a/M \approx 1$, Eqs. (0.26) and (0.49) imply that $P \sim \frac{1}{2}F^{\mu\nu}F_{\nu\mu} \sim (r - r_+)^{-2}(B^P)^2$, our fiducial pressure function.

As with other MHD simulations, the string formalism cannot represent non MHD effects (reconnection) that can cause this scenario to breakdown very near to the horizon [6]. In order to model flows without magnetically arrested accretion (flows with milder pressure divergences near the horizon) require far more computing power because the GHM effects are produced even closer to the hole where there are even larger gradients in all of the quantities. Our hardware and software are not currently capable of computing for long enough to see the jet. We see the negative energy forming and a jet just beginning, then the errors become too large; ending the simulation.